\documentclass[aps,pre,preprint,showpacs,showkeys,12pt]{revtex4}
\usepackage{setspace}
\begin{document}
\title{Circulation in Blowdown Flows}
\author{J. I. Katz}
\affiliation{MITRE Corp., McLean, Va. 22102}
\altaffiliation{Dept. Physics and McDonnell Center for the Space Sciences \\ 
Washington University, St. Louis, Mo. 63130}
\email{katz@wuphys.wustl.edu}
\date{\today}

\renewcommand{\baselinestretch}{2.0}

\begin{abstract}
The blowdown of high pressure gas in a pressure vessel produces rapid
adiabatic cooling of the gas remaining in the vessel.  The gas near the
wall is warmed by conduction from the wall, producing radial temperature
and density gradients that affect the flow, the mass efflux rate and the
thermodynamic states of both the outflowing and the contained gas.  The
resulting buoyancy-driven flow circulates gas through the vessel and
reduces, but does not eliminate, these gradients.  The purpose of this note
is to estimate when blowdown cooling is rapid enough that the gas in the
pressure vessel is neither isothermal nor isopycnic, though it remains
isobaric.  I define a dimensionless number, the buoyancy circulation number
$BC$, that parametrizes these effects.
\end{abstract}



\maketitle





\renewcommand{\baselinestretch}{2.0}
\section{Introduction}

The process by which a pressure vessel containing high pressure gas is
vented to the outside through a narrow (compared to its diameter) orifice
is called blowdown\footnote{This paper concerns pressure vessels that
initially contain only gas.  Blowdown of vessels that initially contain
liquid or solid that flashes into vapor upon pressure release involves
different phenomena.}.  The rapid reduction in pressure and density within
the vessel produces adiabatic cooling of the gas.  If cooling is sufficient,
gases of high condensation temperature may condense as liquid droplets,
leading to a complex multi-phase flow.  These droplets tend to fall to the
bottom of the vessel, where contact with the warm wall will evaporate them,
but if small enough may instead be entrained in the flowing gas.  The
blowdown rate will then depend on the position of the venting orifice and
the nucleation process that controls the droplet size.

In general, the gas inside the vessel may be regarded as isobaric
because the blowdown time (for an orifice narrow compared to the vessel
size) is much longer than the sound transit time across the vessel.  Near
the wall the gas is warmed by conduction from the wall.  Typically, the
vessel itself has a much greater mass and heat content than the contained
gas, and may be regarded as a constant temperature heat bath.  The thermal
diffusion time across the gas is usually much longer than the blowdown
time, so the temperature variation within the gas may be significant.

Under isobaric conditions the temperature gradient implies a density
gradient.  In the presence of a gravitational field this density gradient
drives a circulatory flow, with gas acquiring heat in a boundary layer at
the wall and gradually mixing with the bulk of the gas.  There is some
similarity to Ekman pumping, but in the blowdown problem there is an
additional imposed time scale, the blowdown time, that has no analogue in
Ekman pumping, while Ekman pumping has a characteristic rotational
velocity that has no analogue in the blowdown problem.

\section{Time Scales}

Here we estimate the relevant time scales and other parameters.  The
circulation time $t_{circ}$ is the characteristic time for gas to flow
through the boundary layer in a (nominally spherical) vessel of radius $r$
at speed $v$:
\begin{equation}
t_{circ} \approx {r \over v}.
\end{equation}
The boundary layer thickness is given by
\begin{equation}
\delta \approx \sqrt{D t_{circ}},
\end{equation}
where $D$ is the thermal diffusivity.  The circulatory speed $v$ in a
gravitational field $g$ (or equivalent acceleration) may be estimated,
approximating the flow as freely accelerated (approximately valid if the
Prandtl number is of order unity, as it is for dilute gases):
\begin{equation}
v \approx \sqrt{g r At},
\end{equation}
where $At$ is an Atwood's number:
\begin{equation}
At \approx {t_{circ} \over t_{cool}},
\end{equation}
where $t_{cool}$ is the adiabatic cooling time.

For unchoked flow of an ideal gas with adiabatic exponent $\gamma$ and sound
speed $c_s$ from a volume $V$ through an orifice of area $A$ the adiabatic
cooling time
\begin{equation}
t_{cool} = {V \over A c_s (\gamma - 1)}.
\end{equation}
If flow through the boundary layer is fast compared to this time ($t_{circ}
\ll t_{cool}$), as will usually be the case, $At$ will be small because the
gas in the interior of the vessel will only cool by a small fraction of
its (absolute) temperature in the time required for gas in the boundary
layer to flow through that layer.

Combining the previous results we estimate the circulation time:
\begin{equation}
t_{circ} \approx \left({r t_{cool} \over g}\right)^{1/3}.
\end{equation}
The time required for the entire volume of contained gas to flow through
the boundary layer is
\begin{equation}
t_{mix} \approx t_{circ} {r \over \delta} \approx {r^{7/6} t_{cool}^{1/6}
\over g^{1/6} D^{1/2}}.
\end{equation}
If $t_{mix} > t_{cool}$ then the temperature gradients are large and it
is not valid to treat the gas as isothermal or isopycnic (constant
density).  

\section{The Blowdown Circulation Number}

To facilitate dimensional insight, we rewrite the preceding equation:
\begin{equation}
t_{mix} \approx t_{cool}^{1/6} t_{char}^{5/6},
\end{equation}
where the characteristic time, a quantity defined only by the properties
of the vessel, gas, and gravity, is
\begin{equation}
t_{char} \equiv {r^{7/5} \over g^{1/5} D^{3/5}}.
\end{equation}
This permits defining a (possibly novel) dimensionless parameter, the
blowdown circulation number:
\begin{equation}
BC \equiv {r^{7/6} \over g^{1/6} D^{1/2} t_{cool}^{5/6}} \approx {t_{mix}
\over t_{cool}} \approx \left({t_{char} \over t_{cool}}\right)^{5/6}.
\end{equation}

If $BC \gg 1$ the gas is neither isothermal nor isopycnic and its
circulation and density and temperature distributions must be considered in
order to calculate the blowdown correctly, even if droplets do not condense.
The escaping gas is drawn from near the surface of the pressure vessel,
where it is warmed by thermal conduction from the wall, while the
temperature in the central regions of the vessel drops nearly
adiabatically.  Because $t_{mix}$ is much greater than the actual blowdown
time $t_{cool} (\gamma - 1)$, the gas issuing from the orifice late in the
blowdown process will be much colder than its initial temperature.  A
quantitative calculation would, in general, require a three-dimensional
numerical solution of the Navier-Stokes equations with heat flow.   In the
special case of an orifice at the top or bottom of the pressure vessel the
flow would be two-dimensional with axial symmetry.

\section{Parameters of Circulatory Flow}

Explicitly, the flow parameters $v$, $\delta$ and boundary layer
Reynolds number $Re$ are (for Prandtl number of order unity):
\begin{eqnarray}
v & \approx &{g^{1/3} r^{2/3} \over t_{cool}^{1/3}}; \\
\delta & \approx &{D^{1/2} r^{1/6} t_{cool}^{1/6} \over g^{1/6}};\\
Re & \approx &{g^{1/6} r^{5/6} \over t_{cool}^{1/6} D^{1/2}}.
\end{eqnarray}

For $Re > 10^3$--$10^4$ the boundary layer flow will be turbulent and
$D$ should be replaced by a turbulent diffusion coefficient.  Its value
may be such as to reduce $Re$ to the threshold of turbulent breakdown.
Such large values of $Re$ are only found for very large pressure vessels.

As a numerical example, consider a spherical pressure vessel with $r = 1$ m
with orifice $A = 10^{-3}$ m$^2$ (10 cm$^2$) containing air at a pressure of
$10^7$ Pa (100 bar) and 20$^{\,\circ}$C.  Then $D = 2.12 \times 10^{-7}$
m$^2$/sec (0.00212 cm$^2$/sec), $t_{cool} = 31$ sec, $v = 0.68$ m/sec (68
cm/sec), $\delta = 5.6 \times 10^{-4}$ m (0.056 cm) and $BC = 85$.  The
time for gas to mix throughout the vessel $t_{mix} = BC t_{cool} = 2600$
sec, much longer than the venting time $V/(A c_s) = 12$ sec.  The
temperature and density distributions within the vessel will be strongly
inhomogeneous, even though the pressure remains uniform to high accuracy
everywhere except in the immediate vicinity of the orifice, and, in general,
three-dimensional computational fluid dynamics would be required for a
quantitative calculation of the efflux rate and thermodynamic parameters.

\end{document}